\documentclass[11pt]{article}
\usepackage{color}
\usepackage{amsmath}
\usepackage{amssymb}
\usepackage{amsthm}
\usepackage{epsfig}
\usepackage{tikz}
\usepackage{geometry}
\usepackage{longtable}
\usepackage{supertabular}
\usepackage{url}
\usepackage{blkarray}
\usepackage{enumitem}
\usepackage{setspace}
\usepackage{algorithmic}
\usepackage{algorithm}

\definecolor{wisconsin-red}{rgb}{0.6,0,0}
\newtheorem{proposition}{Proposition}

\newtheorem*{remark}{Remark}

\newcommand{\E}{\mathrm{E}}

\newcommand{\Reals}{\mathbb{R}}

\newcommand{\exclude}[1]{}
\newcommand{\Y}{\mathbf{Y}}
\newcommand{\R}{\mathbf{R}}
\newcommand{\B}{\mathbf{B}}
\newcommand{\bm}{\mathbf}

\newcommand{\BSigma}{\boldsymbol\Sigma}

\newcommand{\Bmu}{\boldsymbol{\mu}}
\newcommand{\Btheta}{\boldsymbol\theta}
\newcommand{\yx}{\hat{y}_{k_{n+1}}}
\newcommand{\y}{\hat{y}^{n+1}_k}

\begin{document}
\author{Qiong Zhang\thanks{Virginia Commonwealth University, USA, qzhang4@vcu.edu} \and Yongjia
Song\thanks{Virginia Commonwealth University, USA, ysong3@vcu.edu} }

\title{Moment Matching Based Conjugacy Approximation for Bayesian Ranking and Selection} 
\maketitle

\begin{abstract}
We study the conjugacy approximation method in the context of Bayesian ranking and selection with unknown correlations. Under the assumption of normal-inverse-Wishart prior distribution, the posterior distribution remains a normal-inverse-Wishart distribution thanks to the conjugacy property when all alternatives are sampled at each step. However, this conjugacy property no longer holds if only one alternative is sampled at a time, an appropriate setting when there is a limited budget on the number of samples. We propose two new conjugacy approximation methods based on the idea of moment matching. Both of them yield closed-form Bayesian prior updating formulas. This updating formula can then be combined with the knowledge gradient algorithm under the  ``value of information'' framework. We conduct computational experiments to show the superiority of the proposed conjugacy approximation methods, including applications in wind farm placement and computer model calibration.
\end{abstract}

\noindent \textbf{Keywords:}\newline Bayesian learning; ranking and selection; moment matching; approximate conjugacy

\section{Introduction}
In this work, we are concerned about selecting the best among a finite set of alternatives. We consider the scenario where we are given a budget to perform a limited number of measurements to evaluate the performances of these alternatives, before the final selection is made. In many real-world applications, the performances of the alternatives may have an underlying but unknown correlation structure, which could be  exploited to improve learning for the whole set of alternatives while only a small number of measurements are performed. This situation arises in a variety of applications. One such example
is computer model parameter calibration where one aims at selecting parameters that best matches the original physical system. Another example is the optimal wind farm placement \cite{Qu2014}, where one selects a candidate location that has the highest expected wind power output. In these applications, it is usually too costly to first measure all the alternatives multiple times and then select the best according to the estimated expected performances. We need to wisely allocate the measurement budget among these alternatives. 

In the literature, this type of problem has been studied under the methodology known as ranking and selection. The basic idea of ranking and selection is to replicate more on ``promising'' candidates. More specifically, ranking and selection first builds a statistical model that quantifies the decision maker's estimation of the expected performances of the alternatives, and then solves an optimization problem to allocate measurement budget among all alternatives. In the literature, ranking and selection is studied under two different streams. In the frequentists' perspective, ranking and selection is based on the indifferent-zone approach \cite{KimNelson2001,KimNelson2006a,KimNelson2007,HongNelson2009}. From the Bayesian perspective, ranking and selection is studied under the ``value of information'' framework, see, e.g., \cite{Chick2006}, and \cite{PowellRyzhov2012} for overviews of the framework. See 
\cite{ChickFrazier2012}, \cite{XieFrazier2013}, and \cite{Qu2014} for recent development of this approach.

We focus on problems where the performances of different alternatives are likely to be correlated, but such a correlation structure is unknown apriori. If a good approximation of this correlation structure is available, it will help to prevent wasting costly measurements on alternatives that are highly correlated, since one may take advantage of the correlation information to learn about other alternatives using measurement results from a single alternative. Classical ranking and selection methods are well-developed for cases where the performances of different alternatives are assumed to be independent (e.g., \cite{PowellRyzhov2012}). Recently, approaches that exploit the underlying correlation structure have been developed. \cite{Frazier2009} study Bayesian ranking and selection for correlated normal beliefs. \cite{Scott2011} build a Gaussian process model to incorporate the correlation information, and their numerical studies show that the correlation matrix can be accurately approximated by a parametric model (based on kernel function or other known structures). However, in many situations it is a luxury to obtain such an accurate approximation of correlation matrix, and we may only be able to gradually learn the correlation structure while making more measurements. Along this line, \cite{Qu2014} recently propose a Bayesian sequential learning procedure based on the normal-inverse-Wishart distribution (e.g., \cite{wishartbook}) to address the issue of unknown correlation matrix. The normal-inverst-Wishart distribution provides a very convenient way of updating the prior distribution (i.e., beliefs about the alternatives) using a simple closed-form updating formula, a property known as ``conjugacy'' in Bayesian statistics. The full conjugacy condition requires that all alternatives should be sampled simultaneously. However, in the context of fully sequential ranking and selection, if only a single alternative is measured in one step, this conjugacy property no longer holds, i.e., the posterior distribution is no longer normal-inverse-Wishart. %\cite{Chick2001a} address this issue by allocating simulation budget equally to a subset of alternatives. To gain more flexibility to sample any alternative at any time, 
\cite{Qu2014} approximate the posterior distribution as a normal-inverse-Wishart distribution by minimizing their Kullback-Leibler divergence. This conjugacy approximation approach still gives rise to a closed-form prior updating formula. They provide extensive computational results to show the superiority of their method compared to many existing methods.
We follow this idea of conjugacy approximation proposed by \cite{Qu2014} and propose alternative approximation methods.

Specifically, we propose a different approximation scheme to match the posterior distribution with a normal-inverse-Wishart distribution, using the idea of matching their first moments. This different approximation scheme is motivated by the fact that matching two distributions by minimizing their Kullback-Leibler divergence, a distance measure of two distributions over all the moments, may be unnecessarily strong and induce some over-fitting issue. In contrast, the parameters required in the updating formula only involve the first and second order moments. Therefore, a complete matching of two distributions over all moments may not be necessary. Along this line, we develop two moment matching based conjugacy approximation for sequential ranking and selection under a normal-inverse-Wishart Bayesian model. 

The contribution of this paper is two-folds. From the methodology perspective, we provide two new alternative conjugacy approximation methods for Bayesian ranking and selection under a normal-inverse-Wishart Bayesian model, both of which also yield closed-form prior updating formulas. We also show that they are superior to the Kullback-Leibler based approximation in \cite{Qu2014} in certain cases according to our numerical study. From the application perspective, this paper is the first one that applies the methodology of Bayesian ranking and selection to calibration of computer models. \exclude{Recently, \cite{1641} develop a consistency theory for approximate Bayesian learning models using a modified version of our results.}

The rest of the paper is organized as follows. In Section 2, we review the normal-inverse-Wishart Bayesian model for sequential learning, and the idea of approximating conjugacy for the Bayesian framework proposed by \cite{Qu2014}. In Section 3, we propose two new methods for updating the prior information in the Bayesian framework. In Section 4, we briefly review the knowledge gradient method used to select the alternative to sample at each step based on the value of information. We show our computational experiment results in Section 5. Proofs of theoretical results are deferred in the appendix.

An extended abstract of this paper appeared in a conference proceeding \cite{zhang15}. This full version of the paper presents an additional conjugacy approximation method that combines the ideas of moment matching and Kullback-Leibler divergence. We also present additional computational experiments motivated by applications in wind farm placement and computer model calibration.

\section{Problem Setup}
We aim to select the best alternative from a candidate set $\{1,\ldots, K\}$ according to their performances. For example, in the wind  farm placement application, we choose the location with the highest wind power; in  the computer experiments calibration, we choose the parameter setting which best matches the physical system. To be specific, 
let $\mu=(\mu_1,\ldots,\mu_K)^\top$ be their true performances, our goal is to find
\[
k^\ast\in\mathrm{argmax}^K_{i=1}\mu_k.
\]
However, $\mu_k$'s are unknown, so we can only choose the best alternative based on our belief about $\mu$. Following the standard assumptions in Bayesian ranking and selection \cite{Frazier2009,Qu2014}, we assume that our belief about $\mu$ follows a multivariate normal distribution (note that $\mu$ is used to denote both true performance and our belief.)
\begin{equation}\label{eq:prior}
\Bmu | \BSigma \sim N_K(\Btheta^0, (q^0)^{-1}\BSigma),\ \BSigma\sim IW_K(\B^0,b^0),
\end{equation} 
where given $\BSigma$, the conditional distribution of $\Bmu$ is a multivariate normal distribution with mean vector $\Btheta^0$ and covariance matrix $(q^0)^{-1}{\BSigma}$,
and $\BSigma$ follows an inverse-Wishart distribution with parameter $\B^0$ and degree of freedom $b^0$. In the literature of Bayesian statistics, the joint distribution of $\Bmu$ and $\BSigma$ is also referred to as the normal-inverse-Wishart distribution. 
The expectation of $\BSigma$ is $\B^0/(b^0-K-1)$, which quantifies the correlation between the performances of different alternatives.

We update our belief based on random measurements of the performances. Let $\hat\Y=(y_1,\ldots, y_K)^\top$ be a sample of the random measurement, which follows a multivariate distribution 
\begin{equation}\label{eq:y}
\hat\Y\sim N_K(\Bmu,\BSigma).
\end{equation}
Our belief about $\mu$ can be updated sequentially as samples $\hat{\Y}_1,\hat{\Y}_2,\ldots $ are collected in a sequence.
The Bayesian sequential selection is very efficient using 
the Bayesian model in \eqref{eq:prior} and \eqref{eq:y}. This is due to the conjugacy property of the normal-inverse-Wishart distribution in \eqref{eq:prior}, which allows us to update the prior information after each new sample in a computationally tractable way \cite{optstat}. Specifically, suppose that the parameters  in \eqref{eq:prior} and \eqref{eq:y} have been updated to $\Btheta^n$, $\B^n$, $q^n$ and $b^n$ at the $n$-th step, i.e., 
\begin{equation}\label{eq:priorn}
\Bmu | \BSigma\sim N_K(\Btheta^n, (q^n)^{-1}\BSigma),\ \BSigma| \hat{\Y}^n\sim IW_K(\B^n,b^n).
\end{equation} 
Given a new sample $\hat{\Y}^{n+1}$, the posterior density function of $\Bmu$ and $\BSigma$ can be computed by combining the density functions of $\hat{\Y}^{n+1}| \Bmu, \BSigma$, $\Bmu|\BSigma$,
and $\BSigma$:
\begin{equation}\label{eq:posterior}
p^{n+1}(\Bmu, \BSigma | \hat{\Y}^{n+1}) \propto p^{n}(\hat{\Y}^{n+1}| \Bmu, \BSigma) p^{n}(\Bmu|\BSigma)p^{n}(\BSigma),
\end{equation}
where
\begin{equation}
p^{n}(\hat{\Y}^{n+1}| \Bmu, \BSigma) \propto |\BSigma|^{-\frac{1}{2}} \exp \Big\{-\frac{q^n}{2}(\hat{\Y}^{n+1}-\Bmu)^\top \BSigma^{-1} (\hat{\Y}^{n+1}-\Bmu) \Big\},
\end{equation}
\begin{equation}
p^n(\Bmu|\BSigma) \propto |\BSigma|^{-\frac{1}{2}}\exp \Big\{-\frac{q^n}{2}(\Bmu-\Btheta^n)^\top \BSigma^{-1} (\Bmu-\Btheta^n)\Big\},
\end{equation}
and
\begin{equation}\label{wishart}
p^n(\BSigma) \propto |\BSigma|^{-\frac{b^n+K+1}{2}}\exp \Big\{-\frac{1}{2}tr(\B^n \BSigma^{-1})\Big\}.
\end{equation}
By combining the above terms, it can be shown that $p^n(\Bmu, \BSigma | \hat{\Y}^{n+1})$ follows a normal-inverse-Wishart distribution with parameters
\begin{eqnarray}\label{all-update}
\nonumber
q^{n+1}& =& q^n + 1 \\\nonumber
b^{n+1}& = &b^n + 1 \\\nonumber
\Btheta^{n+1} & =&\frac{q^n \Btheta^n+ \hat{\Y}^{n+1}}{q^n+1} \\
\B^{n+1} & = &\B^n+ \frac{q^n}{q^n+1}(\Btheta^n-\hat{\Y}^{n+1})(\Btheta^n-\hat{\Y}^{n+1})^\top.  
\end{eqnarray}

The normal-inverse-Wishart distribution provides a very convenient way to update the prior. However, this update requires a sample of all alternatives $\hat\Y$ at each step, which could be too expensive when the number of alternatives is large or sampling is costly. \cite{PowellRyzhov2012} and \cite{Frazier2009} show that it is computationally advantageous to choose the most promising alternative to sample at each step. However, the flexibility of choosing only one alternative at a time will cause a significant challenge: the updating formula \eqref{all-update} cannot be applied in this case. The reason is that
$p^{n+1}(\Bmu, \BSigma | \y)$ ($\neq p^{n+1}(\Bmu, \BSigma |\hat{\Y}^{n+1})$) no longer follows a normal-inverse-Wishart distribution. To address this challenge, two important questions need to be answered: first, how to update the prior information in \eqref{eq:prior} when only alternative is sampled at each step; and second, how to choose the most ``promising'' alternative at each step. For the first question, we review an existing method in the rest of this section, and propose two new methods in Section 3. For the second question, we show in Section 4 how the proposed new methods can be used in the knowledge gradient algorithm \cite{Frazier2009}, where the alternative to sample at each step is chosen by maximizing the value of information.

We now review an existing prior updating method proposed by \cite{Qu2014} for the Bayesian model in \eqref{eq:prior}-\eqref{eq:y} using the idea of approximate conjugacy. For the convenience of presentation, we introduce notations that will be used throughout the rest of the paper.

\paragraph{Notation}\label{note:matrix} For any vector $\mathbf x \in \Reals^{K}$, we denote the $k$-th element of $\mathbf x$ as $\mathbf x_k$, and we denote the vector consisting of all elements of $\mathbf x$ except $\mathbf x_k$ as $\mathbf x_{-k} \in \Reals^{K-1}$. For any $K\times K$ symmetric matrix $\bm X$, we let $\bm X_{kk}$ be the $k$-th diagonal element of $\mathbf X$, $\mathbf X_{\cdot, k}$ be the $k$-th column of $\mathbf X$, $\mathbf X_{-k, k}\in \Reals^{K-1}$ be the subvector of $\mathbf X_{\cdot, k}$ whose $k$-th element is excluded, and $\mathbf X_{-k,-k}$ be the submatrix of $\mathbf X$ constructed by removing the $k$-th row and the $k$-th column of $\mathbf X$. We also define:
\begin{equation*}
\mathbf X_{-k|k} := \mathbf X_{-k,-k} - \frac{\mathbf X_{-k,k}\mathbf X_{k,-k}}{\mathbf X_{kk}}.
\end{equation*}

We first consider how to update the prior information from the $n$-th step, given that 
$k$ is the alternative chosen to be sampled in the $(n+1)$-th step. Given $\Bmu$ and $\BSigma$, 
the new update $\y$ 
follows a normal distribution, $\y \sim N(\Bmu_{k}, \BSigma_{kk})$.
Using the Bayes' rule, the posterior distribution of $\Bmu$ and $\BSigma$ given $\y$ is: 
\begin{equation}\label{eq:densityy}
\begin{split}
p^{n+1}(\Bmu, \BSigma | \y) \propto & |\BSigma|^{-\frac{b^n+K+1}{2}}\exp \Big\{-\frac{1}{2}tr(\B^n \BSigma^{-1})\Big\} 
\\&\cdot |\BSigma|^{-\frac{1}{2}}  \exp\Big\{-\frac{q^n}{2}(\Bmu-\Btheta^n)^\top \BSigma^{-1} (\Bmu-\Btheta^n)\Big\} 
\\ & \cdot \BSigma^{-1/2}_{kk} \exp\Big\{-\frac{(\y-\Bmu_k)^2}{2\BSigma_{kk}} \Big\}.
\end{split}
\end{equation}
We see that the posterior distribution is no longer a normal-inverse-Wishart distribution. Therefore, the conjugacy property of normal-inverse-Wishart distribution cannot be applied. 

To address this issue, \cite{Qu2014} proposed to use the ``optimal approximation of conjugacy'' based on minimizing the Kullback-Leibler divergence between the posterior distribution \eqref{eq:densityy} and a normal-inverse-Wishart distribution, which also leads to a closed-form updating formula as follows:
\begin{eqnarray}\label{qu-update}
\nonumber
q^{n+1}& =& q^{n} + \frac{1}{K} \\\nonumber
b^{n+1}& = &b^{n} + \Delta b^n \\\nonumber
\Btheta^{n+1} & =&\Btheta^n+\frac{\y-\theta^n_{k}}{\frac{b^{n+1}(q^n+1)-K+1}{b^{n+1}-K+1}\B^n_{kk}}\B^n_{\cdot,k} \\\nonumber
\B^{n+1}& = &\frac{b^{n+1}}{b^n}\B^n
+ \frac{b^{n+1}}{b^n+1}\left(\frac{q^n(b^{n+1}-K+1)(\y-\Btheta^n_{k})^2}{b^{n+1}(q^n+1)-K+1}-\frac{\B^n_{kk}}{b^n}\right)\\
&\cdot&\frac{\B^n_{\cdot,k}\B^n_{k,\cdot}}{\B^2_{kk}}
\end{eqnarray}
where $\Delta b^n$ is a number that can be numerically computed by a bisection algorithm, or approximated by $K^{-1}$ according to \cite{Qu2014}. 

Although this framework works well in the numerical experiments shown by \cite{Qu2014}, matching two distributions using Kullback-Leibler divergence is a very strong requirement. The Kullback-Leibler divergence of two distributions is equivalent to a distance measure of two distributions over the moments of all orders. When the true distribution is far away from normal-inverse-Wishart distribution, it may generate over-fitting issues. Therefore, a complete matching of two distributions may not necessarily lead to more accurate approximation. To address this issue, we propose two alternative methods to match the posterior distribution with a normal-inverse-Wishart distribution.

\section{Moment Matching based Approximate Conjugacy}
In this section, we consider two alternative methods to approximate the posterior distribution \eqref{eq:densityy} to a normal-inverse-Wishart distribution using the idea of moment matching. The first method employs the first-order moment matching, and the second method combines the idea of moment matching and Kullback-Leibler divergence minimization. Same as the method in \cite{Qu2014}, both our new proposed methods yield  closed-form updating formulas, which make the Bayesian sequential ranking and selection procedure computationally tractable. A preliminary version of the first approximation method has appeared in a conference proceeding \cite{zhang15}. 

\subsection{Conjugacy approximation based on first-order moment matching}\label{sec:moment}
We consider how to update the prior information in \eqref{eq:prior} in each step given a new observation $\hat{y}^{n+1}_k$. Following \cite{Qu2014}, we set the increase of number of samples as $K^{-1}$ at each step, since only one among $K$ alternatives is sampled. Therefore, we update
$q^{n+1}$ and $b^{n+1}$ by
$q^{n+1}=q^n+K^{-1}$
and $b^{n+1}=b^n+K^{-1}$.

We now consider how to update $\Btheta^{n+1}$ and $\B^{n+1}$. Let us first recall how this is done when we obtain a sample of all alternatives $\hat\Y^{n+1}$ at the $n$th step. Notice that, $p^{n+1}(\Bmu, \BSigma | \hat{\Y}^{n+1})$ in \eqref{eq:posterior} matches the density function of a normal-inverse-Wishart distribution with parameters $q^{n+1}$, $b^{n+1}$, $\Btheta^{n+1}$ and $\B^{n+1}$ in \eqref{all-update}. Meanwhile, this implies that
\begin{equation}\label{eq:alltheta}
\Btheta^{n+1}=\E(\mu|\hat\Y^{n+1})
\end{equation}
and
\begin{equation}\label{eq:allB}
\B^{n+1}=(b^{n+1}-K-1)\mathrm{E}\left\{\Sigma|\hat\Y^{n+1}\right\}=
(b^{n+1}-K-1)\mathrm{E}\left\{q^{n+1}\mathrm{Var}(\mu|\Sigma,\hat\Y^{n+1})|\hat\Y^{n+1}\right\}.
\end{equation}
That is, the updated parameters $\Btheta^{n+1}$ and $\B^{n+1}$ in \eqref{all-update} match the first-order posterior moments of $\mu$ and $q^{n+1}\mathrm{Var}(\mu|\Sigma,\hat\Y^{n+1})$. When only one alternative $k$ is sampled, the posterior distribution $p^{n+1}(\Bmu, \BSigma | \y)$ does not follow a normal-inverse-Wishart distribution. The updating formula \eqref{qu-update} in \cite{Qu2014} is developed by minimizing the Kullback-Leibler divergence between $p^{n+1}(\Bmu, \BSigma | \y)$ and the density function of a normal-inverse-Wishart distribution. Instead of matching the density functions, we develop updating formulas by matching the first-order posterior moments as in \eqref{eq:alltheta} and \eqref{eq:allB}. To do this, we need to compute the posterior moments with regard to $p^{n+1}(\Bmu, \BSigma | \y)$. In Proposition \ref{prop:cond-expect-var} (a) and (b), we decompose $p^{n+1}(\Bmu, \BSigma | \y)$ into a few parts whose first-order moments can be obtained easily. They will then be used to calculate the first-order posterior moments in \eqref{eq:alltheta} and \eqref{eq:allB}.

\begin{proposition}\label{prop:cond-expect-var}
\begin{itemize}
\item[(a)] Given $\BSigma$ and $\y$, $\Bmu$ follows a multivariate normal distribution 
\begin{equation}\label{eq:post-mu}
\Bmu|\BSigma,\y\sim N_K(\tilde\Btheta, (q^{n+1})^{-1}\tilde\BSigma),
\end{equation}
where 
\[
\tilde\Btheta=\Btheta^n +\frac{(\y-\theta^n_{k})\BSigma_{\cdot,k}}{(q^n+1)\BSigma_{kk}},
\]
and
\[
\tilde\BSigma=\frac{q^{n+1}}{q^{n}+1}\left\{\begin{array}{cc}
\frac{q^n+1}{q^n}\BSigma_{-k|k}+\frac{\Sigma_{-k,k}\Sigma_{k,-k}}{\BSigma_{k,k}} & \BSigma_{-k,k}\\
\BSigma_{k,-k} & \BSigma_{kk}\\ 
\end{array}
\right\}
\]
\item[(b)] Let 
$A=\tilde\BSigma_{-k|k}$,
$a=\tilde\BSigma^{-1}_{k,k}\tilde\BSigma_{-k,k}$,
$\tilde a=\tilde\BSigma_{-k,k}$,
and
$c=\tilde\BSigma_{k,k}$,
we have
\begin{equation}\label{eq:a}
a| A,\y\sim N_{K-1}\left(\frac{\B^n_{-k,k}}{\B^n_{k,k}}, \frac{q^n A}{q^{n+1} \B^n_{k,k}} \right)
\end{equation}
\begin{equation}\label{eq:ta}
\tilde a| A,c,\y\sim N_{K-1}\left(\frac{c\B^n_{-k,k}}{\B^n_{k,k}}, \frac{q^n c^2 A }{q^{n+1} B^n_{k,k}} \right)
\end{equation}
\begin{equation}
A|\y\sim IW_{K-1}\left(b^n,\frac{q^{n+1}}{q^n}\B^n_{-k|k}\right)
\end{equation}
\begin{equation}
c|\y\sim IW_{1}\left(b^n-K+2, \frac{q^{n+1}}{(q^{n}+1)}\left[\B^n_{kk}+\frac{q^n}{q^n+1}(\y-\Btheta^n_k)^2\right]\right),
\end{equation}
and $A$ and $c$ are independent.
\end{itemize}
\end{proposition}

As mentioned earlier, we update $\Btheta^{n+1}$ and $\B^{n+1}$ to be
\[
\Btheta^{n+1}=\E(\Bmu|\y),
\]
and 
\[
\B^{n+1}=(b^{n+1}-K-1)\E\left\{q^{n+1}\mathrm{Var}(\mu|\Sigma,\hat\Y^{n+1})|\y\right\}=(b^{n+1}-K-1)\E\left\{\tilde{\Sigma}|\y\right\}.
\]
The expectations can be calculated using the distributions given in Proposition \ref{prop:cond-expect-var}. Proposition \ref{prop-2} summarizes the results.

\begin{proposition}\label{prop-2}
Given $q^{n+1}$ and
$b^{n+1}$,
 the updating formulas of $\Btheta^{n+1}$
 and $\B^{n+1}$ based on moment matching are given by:
\begin{equation}\label{eq:theta}
\Btheta^{n+1} = \Btheta^n + \frac{\B^n_{\cdot,k}}{\B^n_{kk}}\frac{\y-\theta^n_k}{q^n+1},
\end{equation}
\begin{equation}
\B^{n+1}_{-k,-k}=\frac{q^{n+1}(b^{n+1}-K-1)}{b^n-K}\left\{\frac{\B^n_{-k|k}
}{q^n}
+\frac{\tilde q}{q^{n}+1}\left[\frac{\B^n_{-k|k}}{b^n-K}+\frac{\B^n_{-k,k}\B^n_{k,-k}}{\B^n_{kk}}\right]\right\},
\end{equation}
\begin{equation}
\B^{n+1}_{-k,k}=\frac{q^{n+1}(b^{n+1}-K-1)\tilde q}{(q^{n}+1)(b^n-K)}\B^n_{-k,k},
\end{equation}
and
\begin{equation}
\B^{n+1}_{kk}=\frac{q^{n+1}(b^{n+1}-K-1)\tilde q}{(q^{n}+1)(b^n-K)}\B^n_{kk},
\end{equation}
where 
\[
\tilde q=\left[1+\frac{q^n(\yx-\Btheta^n_k)^2}{(q^n+1)\B^n_{kk}}\right].
\]

\end{proposition}

\begin{remark}\label{rm:twofolds}
As shown in Proposition \ref{prop:cond-expect-var}--\ref{prop-2}, the moment matching method contains two folds of moment matching. In the first fold, we match
\begin{equation}\label{eq:firstfoldmean}
\tilde{\mu}=\E(\Bmu|\BSigma,\y)
\end{equation}
and
\begin{equation}\label{eq:firstfoldvar}
\tilde{\Sigma}=q^{n+1}\mathrm{Var}(\Bmu|\BSigma,\y).
\end{equation}
In the second fold, we set $\Btheta^{n+1} = \E(\tilde{\mu}|\y)$ and $\B^{n+1} = (b^{n+1}-K-1)\E(\tilde{\Sigma}|\y)$.
According to \eqref{eq:post-mu}, the moment matching in the first fold also guarantees that the distribution of $\Bmu|\BSigma,\y$ exactly matches a multivariate normal distribution with parameters $\tilde{\mu}$ and $\tilde{\Sigma}$. However, the moment matching in the second fold does not exactly match two distributions.
\end{remark}

Indicated in Remark \ref{rm:twofolds}, the moment matching in the second fold does not exactly match two distributions. We next consider an alternative conjugacy approximation by combining the ideas of moment matching and Kullback-Leibler divergence minimization. Specifically, we use moment matching in the first fold of approximation (which is exact), but use Kullback-Leibler divergence minimization in the second fold.

%\begin{proposition}\label{prop-4} 
%The degree of freedoms $q^{n+1}$ and $b^{n+1}$ will be update by
%$q^{n+1}=q^n+\Delta^n$ and $b^{n+1}=b^n+\Delta^n$ with
%\begin{equation}
%\Delta^n=K^{-1}\sum^K_{i=1}\frac{(\B^n_{i,k})^2}{\B^n_{kk},\B^n_{i,i}}.
%\end{equation}
%\end{proposition}

\subsection{Conjugacy approximation by combining moment matching and Kullback-Leibler divergence minimization}
We now present an alternative conjugacy approximation method that combines the idea of moment matching and minimization of the Kullback-Leibler divergence. According to the proof of Proposition \ref{prop:cond-expect-var} (available in the appendix), the posterior distribution $p^{n+1}(\Bmu,\BSigma|\y)$ can be decomposed to
\[
p^{n+1}(\Bmu,\BSigma|\y)\propto p^{n+1}(\Bmu,|\tilde\BSigma,\y)p^{n+1}(\tilde\BSigma |\y).
\]
Since $\Bmu,|\tilde\BSigma,\y$ follows a multivariate normal distribution, the moment matching and distribution matching give same results as indicated in Remark \ref{rm:twofolds}. However, $p^{n+1}(\tilde\BSigma |\y)$ is not the density function of an inverse-Wishart distribution. Unlike the method in Section \ref{sec:moment}, we consider minimizing the Kullback-Leibler divergence between $p^{n+1}(\tilde\BSigma|\y)$ and an inverse-Wishart distribution to find $\B^{n+1}$. The Kullback-Leibler divergence between $p^{n+1}(\tilde\BSigma|\y)$ and the density function of an inverse-Wishart distribution $\xi(\tilde\BSigma)$ with parameter $\B$ and degree of freedom $b^{n+1}$ is given by:
\begin{equation}\label{eq:kl}
D^n_{KL}(\B)= \E_\xi\left\{\log\frac{\xi(\tilde\BSigma)}{p^{n+1}(\tilde\BSigma|\y)}\right\}.
\end{equation}
$\B^{n+1}$ is then obtained by solving $\mathrm{min}_{\B>0} D^n_{KL}(\B)$, which has a closed-form that we show in Proposition \ref{prop3}.

\begin{proposition}\label{prop3}
$\B^{n+1} = \mathrm{argmin}_{\B>0} D^n_{KL}(\B)$, is given by:
\begin{equation}\label{eq:bkk}
\B_{k,k}=\frac{q^{n+1}(b^{n+1}-K+1)\left[\B^n_{k,k}+\frac{q^n}{q^n+1}(\y-\Btheta^n_k)^2\right]}{(b^n+1)(q^{n}+1)},
\end{equation}
\begin{equation}
\B^{n+1}_{-k,k}=\frac{\B_{k,k}\B^n_{-k,k}}{\B^n_{k,k}},
\end{equation}
and
\begin{equation}
\B^{n+1}_{-k,-k}=\frac{b^{n+1}q^{n+1}}{b^{n}q^{n}}\B^n_{-k|k}+\frac{\B_{-k,k}\B_{k,-k}}{\B_{k,k}}.
\end{equation}

\end{proposition}

By combining \eqref{eq:theta} with the formula of $\B^{n+1}$ in Proposition \ref{prop3}, we obtain a new sequential prior updating procedure.

\begin{remark}\label{rm:compare}
When all alternatives $\hat\Y$ are sampled in each step, both moment matching and Kullback-Leibler divergence minimization will lead to the same updating formula as in \eqref{all-update}.  
\end{remark}

In Section 5, we compare the performances of the proposed two new conjugacy approximation methods with the one proposed in \cite{Qu2014} through experiments motivated by two applications. 

\section{Computation of the Value of Information}
In this section, we follow the knowledge gradient framework developed in \cite{Frazier2009}, \cite{PowellRyzhov2012}, and \cite{Qu2014} on how to sequentially select the alternative to sample in each step. According to this framework, suppose we have obtained $n$ samples, the alternative to be sampled in the next step, $k^{n+1}$, is the one that maximizes the value of information \cite{PowellRyzhov2012}:
\begin{equation}\label{eq:v}
V_n(k)=\E\left[\max_{k'=1,2,\ldots, K} \theta^{n+1}_{k'}|k^{n+1}=k\right] - \mathrm{max}_{k'=1,2,\ldots,K}\theta^n_{k'},
\end{equation}
where the expectation is taken with regard to the predictive distribution of $\theta^{n+1}_k$
given all collected data points.

When the updating formula \eqref{qu-update} derived from minimizing the Kullback-Leibler divergence is used, we have:
\begin{equation}\label{thetanplus}
\Btheta^{n+1}=\Btheta^n+\bm s^n(k)T^n,
\end{equation}
where 
\begin{equation}
\bm s^n(k) = \frac{\sqrt{\frac{q^n+1}{q^n(b^n-K+1)}}}{(\frac{q^n b^{n+1}}{b^{n+1}-K+1}+1)\sqrt{\B^n_{kk}}}\B^n_{.,k},
\end{equation} 
and
\begin{equation}\label{eq:tn}
T^n=\frac{\yx-\Btheta^n_k}{\sqrt{\frac{q^n+1}{q^n(b^n-K+1)}\B^n_{kk}}}.
\end{equation}
According to \cite{Qu2014}, the predictive distribution of $T^n$ is a $t$-distribution with degree of freedom $b^n-K+1$. Thus, the expectation in \eqref{eq:v} can be computed using the properties of the $t$-distribution.

Similarly, when the updating formula \eqref{eq:theta} is derived based on moment matching, or the combination between moment matching and Kullback-Leibler divergence minimization as described in Section 3, $\bm s^n(k)$ in \eqref{thetanplus} is defined as:
\[
\bm s^n(k)=\frac{\B^n_{.,k}}{\sqrt{q^n(q^n+1)(b^n-K+1)\B^n_{kk}}},
\]
and $T^n$ is the same as in \eqref{eq:tn}. 
Therefore, we can also use the predictive distribution of $T^n$, i.e., a $t$-distribution with degree of freedom $b^n-K+1$, to compute the expectation in \eqref{eq:v}. 

For all three conjugacy approximation methods, according to the above analysis, the optimization problem that maximizes \eqref{eq:v} can be written as:
\begin{equation}\label{value-of-info}
\max_{k=1,2,\ldots, K} \ \left\{V_n(k)\right\},
\end{equation}
where
\[
V_n(k) := \E\left[\max_{k'=1,2,\ldots, K} (\theta^n_{k'} +s^n_{k'}(k^{n+1})T^n) \mid k^{n+1}=k\right] - \mathrm{max}_{k'=1,2,\ldots,K}\theta^n_{k'},
\]
and $s^n_{k'}(k^{n+1})$ is the $k'$-th element of vector $\bm s^n(k^{n+1})$. A closed-form solution of \eqref{value-of-info} has been provided by \cite{PowellRyzhov2012} and \cite{Qu2014}.

\section{Numerical Experiments}
We present numerical results to compare three approximate conjugacy methods. 
In particular, we compare the performances of the proposed methods and the one proposed in \cite{Qu2014} based on minimizing the Kullback-Leibler divergence using various test cases. The three methods are labeled as:
\begin{itemize}
\item[1.] KL: Minimizing the Kullback-Leibler  divergence as in \cite{Qu2014} (as described in Section 2).
\item[2.] Moment: Matching the first-order moments as described in Section 3.1.
\item[3.] Moment-KL: Combination of moment matching and Kullback-Leibler  divergence minimization as described in Section 3.2.
\end{itemize}

The performances of the three conjugacy approximation methods are compared using their corresponding opportunity costs at each step. As in \cite{Qu2014}, the opportunity cost of each method in step $n$ is defined by 
\begin{equation}\label{opportunity}
C_{n}=\mathrm{max}_{k}\mu_k-\mu_{\mathrm{argmax}_k\Btheta^{n}_k},
\end{equation}
where $\mu_k$ is the true performance of the $k$th alternative, $\Btheta^n_k$ is the posterior mean given by a certain method at step $n$, and $\mu_{\mathrm{argmax}_k\Btheta^{n}_k}$ is the true performance of the best alternative selected by a certain method at the $n$th step. 
 A smaller opportunity cost indicates that the method is more accurate in selecting the best alternative. We would also expect that $C_n$ decreases with $n$.
 For all cases shown below, we replicate the overall procedure 
 $500$ times, and report the average results. 

\subsection{Data generated from a multivariate normal distribution}\label{sec:normal}
We first consider an example where the samples $\hat{\Y}$ are generated from a multivariate normal distribution. We consider nine alternatives, and let their corresponding true mean values be $\frac{1}{9}, \frac{2}{9}, \ldots, 1$, respectively. The true covariance matrix $A$ is given as: $A_{ij} = (-\rho)^{|i-j|}$, and we consider three different values for $\rho$, $0.1$, $0.5$ and $0.9$, which indicate three different levels of correlation strength, low, median and high, respectively. 

\begin{figure}[!ht]
\centering
\includegraphics[scale=0.6]{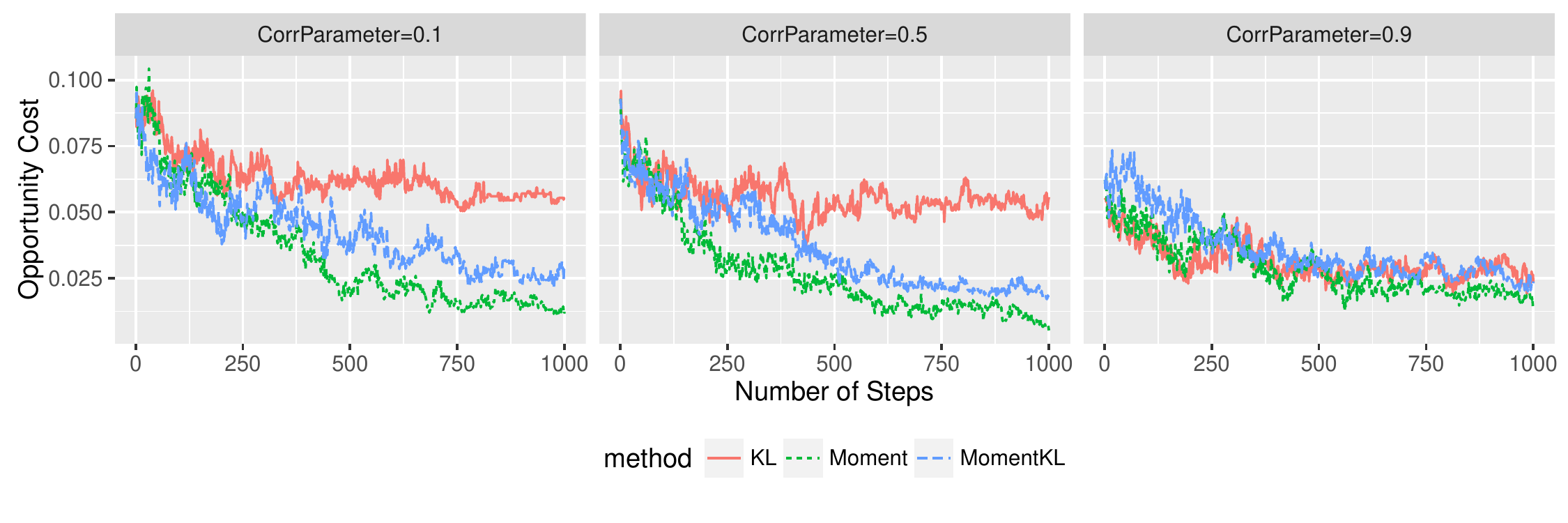}
\caption{Average opportunity cost at each step for $1000$ steps for each method in the multivariate normal case (Section 5.1) over 500 replications.}\label{fg:error}
\end{figure}

Figure \ref{fg:error} shows the performances of the three methods on this example in terms of their opportunity costs \eqref{opportunity} at each step for $1000$ steps. 
The prior parameters are estimated by the sample mean and sample covariance of 
$25$ samples from all alternatives.
We can see from Figure \ref{fg:error} that, as the number of steps increases, the opportunity cost decreases for all three methods. In general, method ``Moment'' gives the smallest opportunity cost for the low and medium correlation cases. However, when the number of steps is small, the performance of method ``Moment-KL'' is comparable with and sometimes better than method ``Moment''. In the high correlation case,  the performances of all three methods are close.

\begin{table}[ht]
\begin{center}
\caption{The mean and standard deviation of the final (at the $1000$-th step) opportunity cost for methods ``KL'', ``Moment'', and ``Moment-KL'' on the multivariate normal example (Section 5.1) with various number of samples for prior estimation, and correlation.}\label{table:error}
\begin{tabular}{ccccccccccc}
\hline
Corr & \# prior & & \multicolumn{2}{c}{KL} & & \multicolumn{2}{c}{Moment} & & \multicolumn{2}{c}{Moment-KL} \\
\hline
 & & & Opp. cost & Error & & Opp. cost & Error & & Opp. cost & Error \\
\cline{4-5} \cline {7-8} \cline{10-11}
0.1 & 5 & & 0.1767 & 0.0023 & & 0.1407 & 0.0023 & & 0.1553 & 0.0021 \\
    & 15 & & 0.0689 & 0.0011 & & 0.0225 & 0.0005 & & 0.0201 & 0.0004 \\
    & 25 & & 0.0554 & 0.0010 & & 0.0123 & 0.0003 & & 0.0286 & 0.0007 \\
\hline 
0.5 & 5 & & 0.1286 & 0.0018 & & 0.1650 & 0.0025 & & 0.1418 & 0.0020 \\
    & 15 & & 0.0844 & 0.0011 & & 0.0149 & 0.0004 & & 0.0195 & 0.0005 \\
    & 25 & & 0.0557 & 0.0009 & & 0.0053 & 0.0002 & & 0.0172 & 0.0005 \\
\hline
0.9 & 5 & & 0.1085 & 0.0017 & & 0.0476 & 0.0008 & & 0.0472 & 0.0010 \\
    & 15 & & 0.0347 & 0.0007 & & 0.0084 & 0.0003 & & 0.0199 & 0.0005 \\
    & 25 & & 0.0233 & 0.0004 & & 0.0149 & 0.0004 & & 0.0238 & 0.0006 \\
\hline
\end{tabular}
\end{center}
\end{table}

We next consider how the performances of three methods vary using different numbers of samples for prior estimation. Table \ref{table:error} shows the means and standard deviations of the final results (at the 1000-th step) on the multivariate normal example with 5, 15, and 25 samples for the prior estimation. In Table \ref{table:error}, columns labeled as ``Opp. cost'' and ``Error'' show the mean and standard deviation of the opportunity cost, respectively (the same labels are also used in Table \ref{table3:error} and Table \ref{table2:error}). We can see from Table \ref{table:error} that, in terms of the final opportunity cost, ``Moment'' and ``Moment-KL'' perform better than ``KL'' in almost all cases considered. 
We also see that, when the prior information is more accurate (when a larger number of samples are used), the opportunity cost is significantly lower for all three methods in most cases.

As observed from both Figure \ref{fg:error} and Table \ref{table:error}, all three methods have similar results when the alternatives are highly correlated. This can be explained by Remark \ref{rm:compare}. When the correlation is high, and the number of alternatives is small (say, $K=9$ in this case), a sample from a single alternative can indicate the performances of other alternatives with high probability. In this sense, sampling a single alternative has a similar effect as sampling all alternatives, in which case the three methods are equivalent as discussed in Remark \ref{rm:compare}. 

\subsection{Wind farm placement using wind speed historical data}
We next study the three methods of conjugacy approximation, ``KL'', ``Moment'', and ``Moment-KL'' on the application of wind farm placement problem using real-world data. 
This application is borrowed from \cite{Qu2014}, where method ``KL'' is compared with several other alternative approaches in the literature. %This paper shows that the ``KL'' based sequential ranking and selection is superior to other methods in the literature.  
In this problem, the goal is to select the best site among a set of candidate sites for installing new wind farms, in terms of average wind power output. We use the publicly available historical wind speed data in the United States from \cite{winddata}.  To be consistent with the results shown in \cite{Qu2014}, we use exactly the same setting described in that paper. However, we may have used a different time period from the wind database \cite{winddata}: we collected hourly wind speed data from June 30th, 2008 to December 31st, 2011, whereas \cite{Qu2014} did not report the range of dates where the data was collected.

\begin{figure}[!ht]
\centering
\includegraphics[scale=0.61]{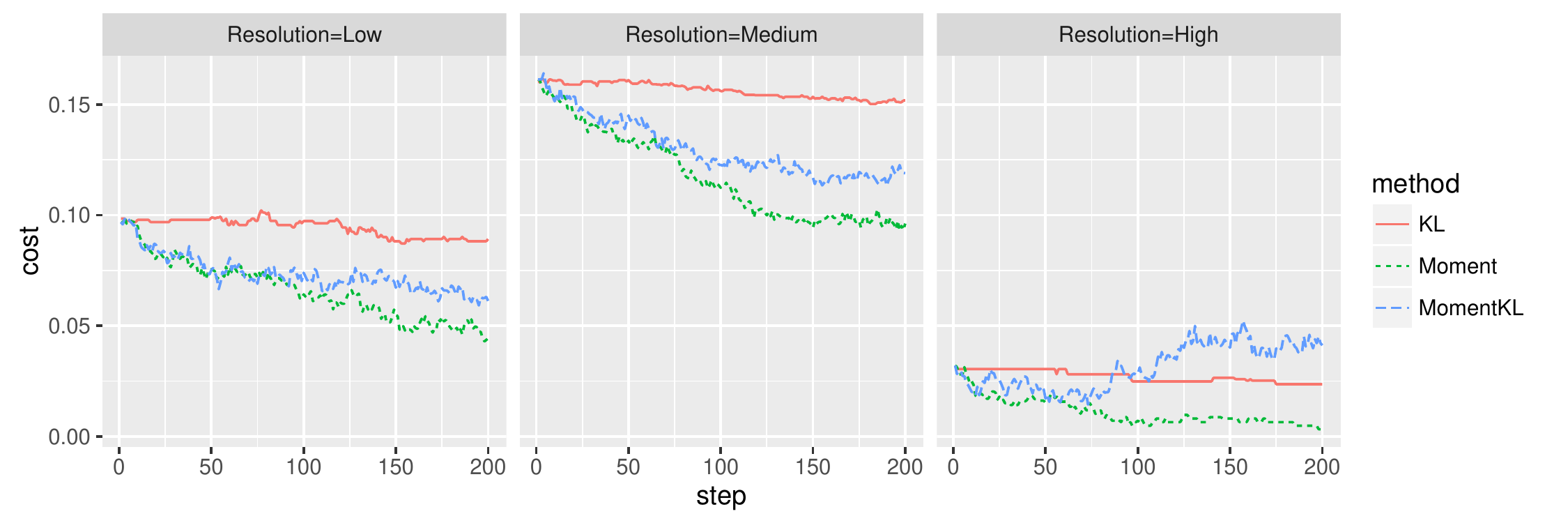}
\caption{Average opportunity cost at each step for $200$ steps for three conjugacy approximation method in the Bayesian sequential ranking and selection in the wind farm placement test case (Section 5.2) over $500$ replications.}\label{fg:wind}
\end{figure}

As in \cite{Qu2014}, we choose from $64$ candidate sites distributed on an $8\times 8$ grid from the state of Washington. We use three different levels of latitude and longitude resolutions, that is, $0.125$ degrees (High), $0.25$ degrees (Medium) and $0.375$ degrees (Low). A higher resolution means more spatial correlations between different locations, and less differences between their true means.
 Figure \ref{fg:wind} shows the average performances of the three conjugacy approximation methods in each step for $200$ steps over $500$ replications. Table \ref{table3:error} shows the average means and standard deviations of the final opportunity cost (at the $200$-th step) of the three methods over $500$ replications. Similar to what we have observed in Section \ref{sec:normal}, the proposed methods ``Moment'' and ``Moment-KL'' perform better than ``KL'' in most scenarios.  Furthermore, we observe in Figure \ref{fg:wind} that the performances under three resolutions are significantly different from each other. This can be explained by the different resolutions considered in the three cases. For the low resolution case (the distance between two alternatives is large), the true performances of different alternatives are significantly different from each other, therefore, it is easy to distinguish among these alternatives, and make the correct selection, which ends up with a small opportunity cost. For the high resolution case (the distance between two alternatives is small), the true performances of different alternatives are similar, therefore, even if a wrong selection is made, it does not lead to a large opportunity cost. The medium resolution case does not enjoy the advantages in either low or high resolution  case, and it gives the worst results in terms of opportunity cost among the three cases.  

\begin{table}[ht]
\begin{center}
\caption{The mean and standard deviation of the final opportunity cost (at the $200$-th step) for methods ``KL'', ``Moment'', and ``Moment-KL'' on the wind farm example (Section 5.2) with different resolutions.} \label{table3:error}
\begin{tabular}{cccccccccc}
\hline
Resolution & & \multicolumn{2}{c}{KL} & & \multicolumn{2}{c}{Moment} & & \multicolumn{2}{c}{Moment-KL} \\
\hline
  & & Opp. cost & Error & & Opp. cost & Error & & Opp. cost & Error \\
\cline{3-4} \cline {6-7} \cline{9-10}
Low & & 0.0892 & 0.0093 & & 0.0443 & 0.0069 & & 0.0613 & 0.0076 \\

Medium & & 0.1518 & 0.0089 & & 0.0962 & 0.0068 & & 0.1190 & 0.0074 \\

High & & 0.0236 & 0.0061 & & 0.0032 & 0.0023 & & 0.0412 & 0.0078 \\
\hline
\end{tabular}
\end{center}
\end{table}

\subsection{Computer model calibration}
In this section, we formulate the computer model calibration problem as a Bayesian ranking and selection problem. Consider a physical system with $\bm x\in \mathcal X\subset\R^d$ being the control variables. The response of the system can be seen as a real-valued stochastic function, denoted by $\eta(\bm x)$. When running physical experiments is costly, a statistical predictor $\hat \eta(\bm x)$ (such as the interpolator in \cite{tuo2015efficient}) can be used to model the unknown true response $\eta(\bm x)$ based on a set of observations. %Due to the unavailability of $\eta(\bm x)$ at most input points, $\hat \eta(\bm x)$ is used as a surrogate of $\eta(\bm x)$ in the rest of this paper.
%{\color{red} In this paper, we assume that $\hat\eta(\bm x)$ is the best predictor under current condition, and we are not able to obtain more observations.}

Computer experiments are usually used to mimic costly physical experiments. 
Input parameters of a computer model include control variables $\bm x$ in the physical system, as well as a calibration variable $\lambda$, which describes some inherent features of the physical system. Let the response of this computer model be
$f(\bm x,\lambda)$, the goal of calibrating this computer experiment is to reduce the gap between $f(\bm x,\lambda)$ and $\eta(\bm x)$ by choosing an appropriate $\lambda$. We consider the case where the calibration variable $\lambda$ is a qualitative parameter with $K$ different qualitative levels. In cases where multiple qualitative variables exist, we let $\lambda$ be an aggregate qualitative parameter whose qualitative levels correspond to all level combinations of these variables. Similar to \cite{tuo2015efficient}, the calibration variable $\lambda$ is chosen by minimizing the mean squared error (MSE) between the physical model and the computer model:
\begin{equation}\label{eq:obj}
\mathrm{MSE}(\lambda)=\E\left\{\eta(\bm x)-f(\bm x, \lambda)\right\}^2,
\end{equation}
where the expectation is taken with regard to the randomness of $\bm x$, and the randomness of the physical system and/or the computer model (depending on whether or not the computer model is stochastic). The mean squared error in \eqref{eq:obj} measures the model discrepancy of $f(\bm x, \lambda)$.
Since $f(\bm x, \lambda)$ and $\eta(\bm x)$ are unknown and only available at a few design points, the MSE in \eqref{eq:obj} is not readily available for each $\lambda$. By surrogating $\eta(\bm x)$ with $\hat\eta(\bm x)$, the MSE of each qualitative level could be estimated empirically. By setting $\lambda$ at its $k$-th qualitative level, we run the computer model on a design of control variables $D=\{\bm x_1,\ldots, \bm x_m\}$, and denote the outputs as $f_k(\bm x_i)$ for $i=1,\ldots, m$. 
The empirical estimation of \eqref{eq:obj} is given by:
\begin{equation}\label{sample}
\hat y_k=m^{-1}\sum^m_{i=1} \left\{\hat \eta(\bm x_i)-f_k(\bm x_i)\right\}^2.
\end{equation}
We arrange $\hat y_k$'s from all qualitative levels in a single vector
\begin{equation}\label{eq:Ydf}
\hat{\Y}=(\hat y_1,\ldots,\hat y_K)^\top,
\end{equation}
which estimates the model discrepancy of $f(\bm x, \lambda)$ at all $K$ qualitative levels of $\lambda$.
Hence, we have formulated this problem into a Bayesian ranking and selection problem: the samples are the estimates of MSEs, and the alternatives are the qualitative levels of the calibration parameter $\lambda$ indexed from $\{1,\ldots, K\}$.

We test the three conjugacy approximation methods, ``KL'', ``Moment'', and ``Moment-KL'', for computer model calibration on the Borehole function \cite{borehole}, a widely used example for illustrating various methods in computer experiments. This function models the flow rate of water through a borehole, and has the following form:
\begin{equation}\label{eq:obore}
f(\bm x)=\log\left\{\frac{2\pi x_1 x_6}{\log(x_5/x_2)\left[1+\frac{2x_3x_1}{\log(x_5/x_2)x^2_2x_7}+\frac{x_1}{x_4}\right]}\right\},
\end{equation}
where $\bm x=(x_1,\ldots, x_7)^\top$, and
 the ranges and units of inputs $x_1\sim x_7$ are given in Table \ref{tb:borehole}. Inputs $x_1$--$x_5$ are the control variables of this system, and inputs $x_6$ and  $x_7$ are the qualitative calibration parameters. Function \eqref{eq:obore} is used as the computer model, and the true physical system is specified as
\begin{equation}\label{eq:truebore}
\hat{\eta}(\bm x)=\log\left\{\frac{2\pi x_1 \times 401}{\log(x_5/x_2)\left[1+\frac{2x_3x_1}{\log(x_5/x_2)x^2_2\times 11000}+\frac{x_1}{x_4}\right]}\right\}+N(0,1).
\end{equation}

\begin{table}[!ht]
\begin{center}
\caption{Ranges of the inputs $x_1\sim x_7$ on the Borehole function.}\label{tb:borehole}
\begin{tabular}{|ccc|ccc|}
\hline
Variable & Range & Unit & Variable & Range & Unit\\\hline
$x_1$    & 63070-115600 & $m^2/yr$ & $x_5$ & 100-50000 & $M$\\
$x_2$    & 0.05-0.15 & $M$ & $x_6$ & 170-410 & $M$\\
$x_3$    & 1120-1680 & $M$ & $x_7$ &9588-12045 & $m/yr$\\
$x_4$    & 63.1-116 & $m^2/yr$ & &&\\\hline
\end{tabular}
\end{center}
\end{table}

In our experiments, we compute $\y$ in each step according to \eqref{sample}, where $\hat{\eta}(\cdot)$ function is given by \eqref{eq:truebore}. We let $x_6$ be a qualitative parameter with three equally spaced levels, and we consider two different numbers of levels for parameter $x_7$, $10$ and $17$, which gives $30$ and $51$ level combinations in total, respectively. For each qualitative level, we generate the design points of the control variables $\bm x$ using a $5$-dimensional Latin hypercube design with eight runs. 

\begin{figure}[!ht]
\centering
\includegraphics[scale=0.5]{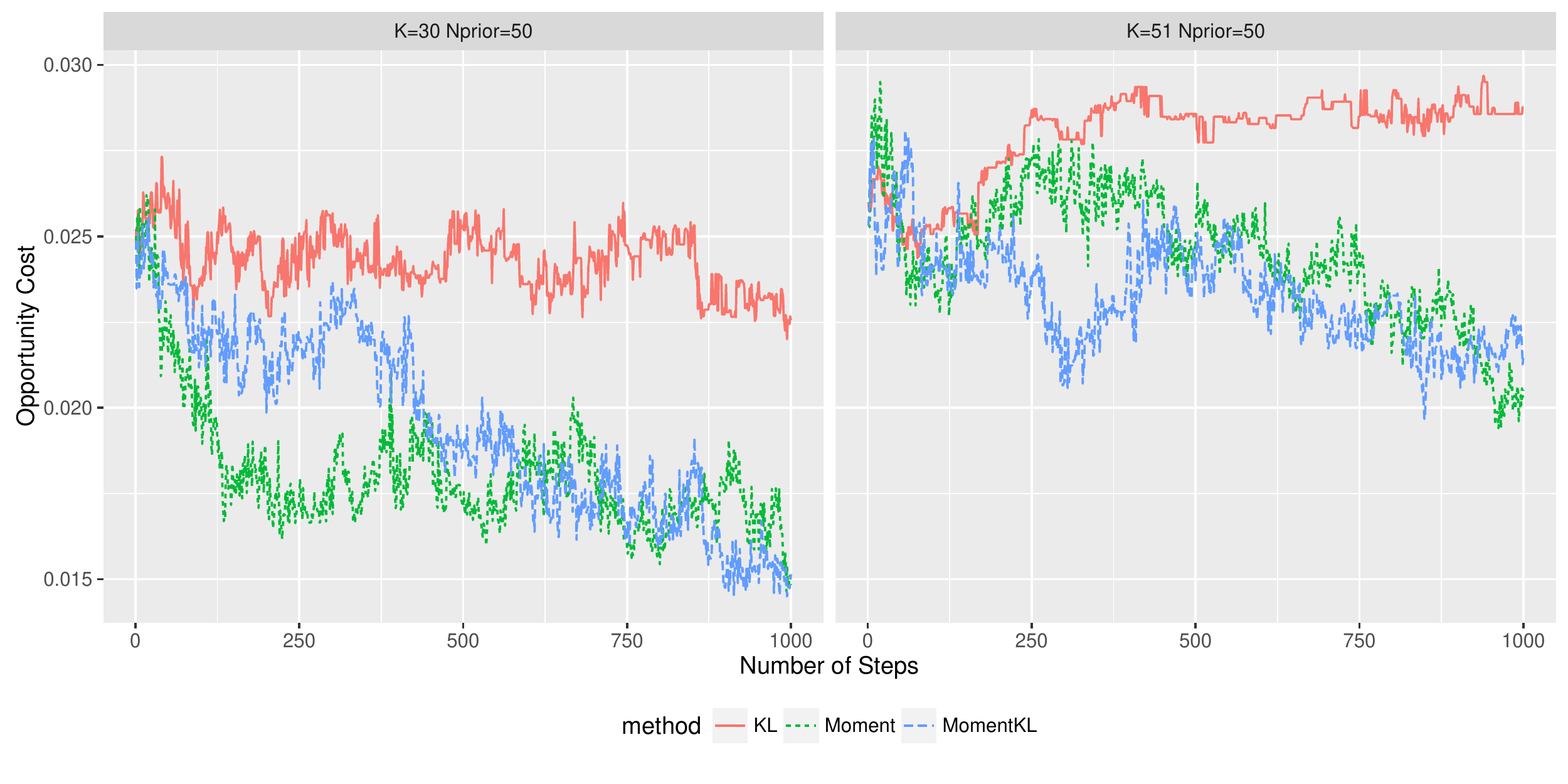}
\caption{Average opportunity cost at each step over $1000$ steps for each methods ``KL'', ``Moment'', and ``Moment-KL'' in the computer model calibration with borehole function (Section 5.3) with $30$ and $51$ qualitative levels over 500 replications.}\label{fg:borehole}
\end{figure}

\begin{table}[ht]
\begin{center}
\caption{The mean and standard deviation of the final opportunity cost (at the $1000$-th step) for methods ``KL'', ``Moment'', and ``Moment-KL'' in the computer model calibration with borehole function (Section 5.3) with various level combinations and number of samples for prior estimation.}\label{table2:error}
\begin{tabular}{ccccccccccc}
\hline
K & \# prior & & \multicolumn{2}{c}{KL} & & \multicolumn{2}{c}{Moment} & & \multicolumn{2}{c}{Moment-KL} \\
\hline
 & & & Opp. cost & Error & & Opp. cost & Error & & Opp. cost & Error \\
\cline{4-5} \cline {7-8} \cline{10-11}
30 & 20 & & 0.0315 & 0.0003 & & 0.0196 & 0.0002 & & 0.0334 & 0.0004 \\
    & 50 & & 0.0226 & 0.0002 & & 0.0148 & 0.0001 & & 0.0151 & 0.0001 \\
\hline 
51 & 20 & & 0.0347 & 0.0004 & & 0.0194 & 0.0002 & & 0.0223 & 0.0003 \\
    & 50 & & 0.0288 & 0.0002 & & 0.0205 & 0.0002 & & 0.0215 & 0.0002 \\
\hline
\end{tabular}
\end{center}
\end{table}

Table \ref{table2:error} and Figure \ref{fg:borehole} show the performances of three different methods. Consistent with what we have observed in the multivariate normal case, method ``Moment'' performs better than the other two methods, especially in the case when the number of qualitative levels is small, and a small number of samples are used to estimate the prior distribution. In cases with larger number of qualitative levels, the performances of the proposed two new conjugacy approximation methods ``Moment'' and ``Moment-KL'' are competitive, and both significantly outperform method ``KL''. We also see that,  when a larger number of samples are used to estimate the prior, the performances of two ``KL'' based methods are significantly improved. However, this is not the case for method ``Moment''. This shows that method ``Moment'' is less sensitive to the accuracy of the prior distribution.

\section{Concluding Remarks}
We have proposed two alternative conjugacy approximation methods for Bayesian ranking and selection. Unlike the distribution match conjugacy approximation in \cite{Qu2014}, our proposal is developed based on moment matching. We have conducted comprehensive numerical experiments, including the applications of the Bayesian ranking and selection on wind farm placement and computer model calibration. Our experiment results have shown the superiority of the proposed methods.

\bibliographystyle{plain}

\medskip

\appendix
\section{Proof of Proposition 1}
\begin{proof}
\noindent{\bf Proof of Proposition 1 (a)}
Given $\BSigma$ and $\yx$, the density function of $\Bmu$ is given by:
\begin{equation}\label{single-post}
p^{n+1}(\Bmu|\BSigma, \yx) \propto  \exp\Big\{-\frac{q^n}{2}(\Bmu-\Btheta^n)^\top \BSigma^{-1} (\Bmu-\Btheta^n)-\frac{(\yx-\mu_k)^2}{2\BSigma_{kk}}\Big\}.
\end{equation}
To show \eqref{single-post} is a multivariate normal distribution, we need to find $\tilde\Btheta$ and $\tilde\BSigma$ that satisfy
\begin{equation}\label{single2}
p^{n+1}(\Bmu|\BSigma, \yx) \propto  \exp\Big\{-\frac{q^{n+1}}{2}(\Bmu-\tilde\Btheta)^\top \tilde\BSigma^{-1} (\Bmu-\tilde\Btheta)\Big\}.
\end{equation}
By comparing \eqref{single-post} and \eqref{single2}, $\tilde\Btheta$ and $(q^{n+1})^{-1}\tilde\BSigma$ should satisfy
\begin{equation}
q^{n+1}\tilde \BSigma^{-1}\tilde\Btheta=q^n \BSigma^{-1}\Btheta^n+\frac{\hat y^{n+1}_k}{\BSigma_{k,k}} e_k
\end{equation}
and 
\begin{equation}
q^{n+1}\tilde\BSigma^{-1}=q^n  \BSigma^{-1}+\frac{1}{\BSigma_{kk}}e_k(e_k)^{\top}
\end{equation}
where $e_k$ is a $K$-dimensional vector whose $k$-th entry equals to 1 and other entries equal to 0. By applying the Sherman-Morrison-Woodbury matrix formula \cite{Woodbury}, 
%\[
%(A-BD^{-1}C)^{-1}=A^{-1}+A^{-1}B(D-CA^{-1}B)^{-1}CA^{-1},
%\]
%with 
%$A=q^n\BSigma$,
%$B=e_k$,
%$C=(e_k)^\top$,
%and
%$D=-\BSigma_{kk}$,
we obtain the formula of 
$\tilde\Btheta$ and $\tilde\BSigma$ 
as in Proposition 1 (a).

\vspace{0.5cm}

\noindent{\bf Proof of Proposition 1 (b)}
Since $A$, $a$, $\tilde a$, and $c$ are functions of $\BSigma$, we first derive the density function of $\BSigma$.
According to the proof in Proposition 1 (a), we have that
\[
p^{n+1}(\Bmu|\BSigma, \yx) \propto  \exp\Big\{-\frac{q^n}{2}(\Bmu-\Btheta^n)^\top \BSigma^{-1} (\Bmu-\Btheta^n)-\frac{(\yx-\mu_k)^2}{2\BSigma_{kk}}\Big\}
\]
\[
= \exp\Big\{-\frac{q^{n+1}}{2}(\Bmu-\tilde\Btheta)^\top \tilde\BSigma^{-1} (\Bmu-\tilde\Btheta)-\frac{q^n(\yx-\theta^n_k)^2}{2(q^n+1)\BSigma_{kk}}\Big\}.
\]
Thus, 
\[
p^{n+1}(\BSigma|\yx)=\int p^{n+1}(\Bmu,\BSigma|\yx)d\Bmu
\]
\[
=  |\BSigma|^{-\frac{b^n+K+2}{2}}\BSigma^{-1/2}_{kk}
\cdot \exp\Big\{-\frac{q^n(\yx-\theta^n_k)^2}{2(q^n+1)\BSigma_{kk}}-\frac{1}{2}\mathrm{tr}(\B^n\BSigma^{-1})\Big\} \]
\[
\cdot\int \exp\Big\{-\frac{q^{n+1}}{2}(\Bmu-\tilde\Btheta)^\top \tilde\BSigma^{-1} (\Bmu-\tilde\Btheta)\Big\}d\Bmu
\]
\begin{equation}\label{eq:sigma}
\propto|\BSigma|^{-\frac{b^n+K+2}{2}}\BSigma^{-1/2}_{kk}
\cdot \exp\Big\{-\frac{q^n(\yx-\theta^n_k)^2}{2(q^n+1)\BSigma_{kk}}-\frac{1}{2}\mathrm{tr}(\B^n\BSigma^{-1})\Big\} 
\cdot |\tilde \BSigma|^{1/2}.
\end{equation}

We now transform the variables in \eqref{eq:sigma} in terms of $A$, $a$, and $c$. Since 
\begin{equation}\label{eq:A}
A=\tilde\BSigma_{-k|k}=\frac{q^{n+1}}{q^n}\BSigma_{-k|k},
\end{equation}
\begin{equation}\label{eq:a}
a=\tilde\BSigma^{-1}_{kk}\tilde\BSigma_{-k,k}=\BSigma^{-1}_{kk}\BSigma_{-k,k},
\end{equation}
and
\begin{equation}\label{eq:c}
c=\tilde \BSigma_{kk}=\frac{q^{n+1}}{q^n+1}\BSigma_{kk},
\end{equation}
we express
\begin{equation}
|\tilde \BSigma|=|\tilde \BSigma_{-k|k}|\cdot \tilde\BSigma_{kk}=|A|\cdot c,
\end{equation}
\begin{equation}
|\BSigma|=|\BSigma_{-k|k}|\cdot \BSigma_{kk}\propto |A|\cdot c,
\end{equation}
and 
\[
\mathrm{tr}(\B^n\BSigma^{-1})=\frac{\B^n_{kk}}{\BSigma_{kk}}+\mathrm{tr}(\B^n_{-k|k}\BSigma^{-1}_{-k|k})+\B^n_{kk}\left (\frac{\BSigma_{-k,k}}{\BSigma_{kk}}-\frac{\B^n_{-k,k}}{\B^n_{kk}}\right)^\top \BSigma^{-1}_{-k|k}\left (\frac{\BSigma_{-k,k}}{\BSigma_{kk}}-\frac{\B^n_{-k,k}}{\B^n_{kk}}\right)
\]
\begin{equation}
=\frac{q^{n+1}\B^n_{kk}}{(q^n+1)c}+\frac{q^{n+1}}{q^n}\mathrm{tr}(\B^n_{-k|k}A^{-1})+\frac{q^{n+1}}{q^n}\B^n_{kk}\left (a-\frac{\B^n_{-k,k}}{\B^n_{kk}}\right)^\top A^{-1}\left (a-\frac{\B^n_{-k,k}}{\B^n_{kk}}\right).
\end{equation}

The determinant of the Jacobin matrix for the transformations in  \eqref{eq:A}--\eqref{eq:c} is $c^{K-1}$. Therefore, we express
\[
p^{n+1}(A,a,c|\yx)\propto |A|^{-\frac{b^n+K+1}{2}} \exp\left\{-\frac{q^{n+1}}{2 q^n}\mathrm{tr}(\B^n_{-k|k}A^{-1})\right\}
\]
\[
\cdot c^{-\frac{b^n-K+4}{2}}\exp\left\{-\frac{q^{n+1}\B^n_{kk}}{2(q^n+1)c}-\frac{q^n q^{n+1}(\yx-\theta^n_k)^2}{2(q^n+1)^2c}\right\}\]
\begin{equation}\label{eq:dist}
\cdot\exp\left\{-\frac{\B^n_{kk}q^{n+1}}{2q^n}\left (a-\frac{\B^n_{-k,k}}{\B^n_{kk}}\right)^\top A^{-1}\left (a-\frac{\B^n_{-k,k}}{\B^n_{kk}}\right)\right\}.
\end{equation}
According to \eqref{eq:dist}, we have
\[
p^{n+1}(a|A, c,\yx)\propto \exp\left\{-\frac{\B^n_{kk}q^{n+1}}{2q^n}\left (a-\frac{\B^n_{-k,k}}{\B^n_{kk}}\right)^\top A^{-1}\left (a-\frac{\B^n_{-k,k}}{\B^n_{kk}}\right)\right\},
\]
which further leads to the multivariate normal distributions of $a$ and $\tilde a$. 
By integrating over $a$ in \eqref{eq:dist}, we have
\[
p^{n+1}(A,c|\yx) \propto |A|^{-\frac{b^n+K}{2}} \exp\left\{-\frac{q^{n+1}}{2 q^n}\mathrm{tr}(\B^n_{-k|k}A^{-1})\right\}
\]
\[
\cdot c^{-\frac{b^n-K+4}{2}}\exp\left\{-\frac{q^{n+1}\B^n_{kk}}{2(q^n+1)c}-\frac{q^n q^{n+1}(\yx-\theta^n_k)^2}{2(q^n+1)^2c}\right\},
\]
which leads to the independent Inverse-Wishart distributions of $A$ and $c$.
\end{proof}

\section{Proof of Proposition 2}
\begin{proof}
We update $\Btheta^{n+1}$ and $\B^{n+1}$ by matching them with the posterior moments of $\tilde\Btheta$ and $\tilde\BSigma$ in Proposition 1, i.e.,
\begin{equation}
\Btheta^{n+1}=\mathrm{E}(\tilde\Btheta|\yx)
\end{equation}
and
\begin{equation}\label{eq:Bn}
\B^{n+1}=(b^{n+1}-K-1)\mathrm{E}(\tilde\BSigma|\yx).
\end{equation}
Therefore, the tasks in this proposition is to derive $\mathrm{E}(\tilde\Btheta|\yx)$ and $\mathrm{E}(\tilde\BSigma|\yx)$.

We first derive $\mathrm{E}(\tilde\Btheta|\yx)$. Recall that
\[
\tilde\Btheta=\Btheta^n +\frac{(\yx-\theta^n_{k})\BSigma_{\cdot,k}}{(q^n+1)\BSigma_{kk}}.
\]
Thus,
\[
\mathrm{E}(\tilde\Btheta|\yx)=\Btheta^n +\frac{(\yx-\theta^n_{k})}{(q^n+1)}\mathrm{E}\frac{\BSigma_{\cdot,k}}{\BSigma_{kk}}.
\]
According to the proof of Proposition 1 (b), $\BSigma_{\cdot,k}/\BSigma_{kk}$ is a vector whose $k$-th component equals to 1, and other components equal to the entries in $a$ defined in Proposition 1. We see from Proposition 1 that, given $\yx$ and $A$, $a$ follows a normal distribution with mean $\B^{n}_{-k,k}/\B^n_{kk}$. Thus, we obtain the expression of $\Btheta^{n+1}$.

Now we derive $\mathrm{E}(\tilde\BSigma|\yx)$. According to the definition of $A$, $a$, $\tilde a$ and $c$ in Proposition 1, we have
\[
\tilde\BSigma_{-k,-k}=A+caa^\top,
\]
\[
\tilde \BSigma_{-k,k}=\tilde a,
\]
and 
\[
\tilde\BSigma_{k,k}=c.
\]
The distributions of $A$, $a$, $\tilde a$ and $c$ are given in Proposition 1 (b). The expectations of $A$, $\tilde a$, and $c$ can be directly given as
\begin{equation}\label{eq:EA}
\mathrm{E}(A|\yx)=\frac{q^{n+1}\B^n_{-k|k}}{q^n (b^n-K)},
\end{equation}
\begin{equation}\label{eq:Ec}
\mathrm{E}(c|\yx)=\frac{q^{n+1}}{(q^{n}+1)(b^n-K)}\left[\B^n_{kk}+\frac{q^n}{q^n+1}(\yx-\Btheta^n_k)^2\right],
\end{equation}
and
\[
\mathrm{E}(\tilde a|\yx)=\mathrm{E}\left\{\mathrm{E}(\tilde a|A,c,\yx)|\yx\right\}=\mathrm{E}(c|\yx)\frac{\B^n_{-k,k}}{\B^n_{kk}}
\]
\begin{equation}\label{eq:Ea}
=\frac{q^{n+1}}{(q^{n}+1)(b^n-K)}\left[1+\frac{q^n (\yx-\Btheta^n_k)^2}{(q^n+1)\B^n_{kk}}\right]\B^n_{-k,k}.
\end{equation}

Now we consider $\mathrm{E}(caa^\top|\yx)$.  According to the proof of Proposition 1(b)
\[
\mathrm{E}(caa^\top|\yx)=\mathrm{E}\{\mathrm{E}(caa^\top|A,c,\yx)|\yx\}
\]
\[
=\mathrm{E}\{\mathrm{E}(caa^\top|A,c,\yx)|\yx\}
\]
\[
=\mathrm{E}\left\{c\left[\mathrm{Var}(a|A,c,\yx)+\mathrm{E}(a|A,c,\yx)\mathrm{E}(a^\top|A,c,\yx)\right]|\yx\right\}
\]
\[
=\mathrm{E}\left\{c \left[\frac{q^n A}{q^{n+1} \B^n_{k,k}}+\frac{\B^n_{-k,k}\B^n_{k,-k}}{(\B^n_{kk})^2}\right]\big|\yx\right\}
\]
\[
=\mathrm{E}(c|\yx)\left[\frac{q^n \mathrm{E}(A|\yx)}{q^{n+1} \B^n_{k,k}}+\frac{\B^n_{-k,k}\B^n_{k,-k}}{(\B^n_{kk})^2}\right]
\]
\begin{equation}\label{eq:Ecaa}
=\frac{q^{n+1}}{(q^{n}+1)(b^n-K)}\left[1+\frac{q^n(\yx-\Btheta^n_k)^2}{(q^n+1)\B^n_{kk}}\right]\left[\frac{\B^n_{-k|k}}{b^n-K}+\frac{\B^n_{-k,k}\B^n_{k,-k}}{\B^n_{kk}}\right]
\end{equation}
Combining the results in \eqref{eq:Bn} and \eqref{eq:A}--\eqref{eq:Ecaa}, we obtain the updating formulas for $\B^{n+1}$.

\end{proof}

\section{Proof of Proposition 3}
\begin{proof}
We decompose the density function of $\tilde\BSigma$ to
\begin{equation}
\xi(\tilde\BSigma)\propto\xi^0\xi(A)\xi(a|A)\xi(c),
\end{equation}
where $A$, $a$ and $c$ are defined in Proposition 1, and
\[
\xi^0= |\B|^{b^{n+1}/2}=|\B_{-k|k}|^{b^{n+1}/2}\cdot \B^{b^{n+1}/2}_{kk},
\]
\[
\xi(a|A)=\exp\left\{-\frac{1}{2}\left(a-\frac{\B_{-k,k}}{\B_{kk}}\right)^\top \left(\frac{A}{\B_{kk}}\right)^{-1}\left(a-\frac{\B_{-k,k}}{\B_{kk}}\right) \right\},
\]
\[
\xi(c)=c^{-\frac{b^{n+1}+K+1}{2}}\exp\left\{-\frac{1}{2}\B_{kk}c^{-1}\right\},
\]
and
\[
\xi(A)=|A|^{-\frac{b^{n+1}+K+1}{2}}\exp\left\{-\frac{1}{2}\mathrm{tr}(\B_{-k|k}A^{-1})\right\}.
\]
According to the properties of the Inverse-Wishart distribution, we have
\begin{equation}
a|A\sim N_{K-1}(\B_{-k,k}/\B_{kk}, A/\B_{kk}),
\end{equation} 
\begin{equation}
c\sim IW_1(\B_{kk}, b^{n+1}-K+1),
\end{equation}
and
\begin{equation}
A\sim IW_{K-1}(\B_{-k|k},b^{n+1}).
\end{equation}

According to Proposition 2, we have that the variance of $\Bmu|\BSigma, \y$ is
$\tilde\BSigma$. According to the proof of Proposition 2, 
the density function of $\tilde\BSigma|\y$ can be decomposed by 
\begin{equation}\label{eq:psig}
p^{n+1}(\tilde\BSigma|\y)\propto p(a|A,\y)p(A|\y)p(c|\y),
\end{equation}
where
\[
p(a|A,\y)=\exp\left\{-\frac{q^{n+1}\B^n_{k,k}}{2q^n}(a-\frac{\B^n_{-k,k}}{\B^n_{k,k}})^\top A^{-1} (a-\frac{\B^n_{-k,k}}{\B^n_{k,k}})\right\},
\]
\[
p(c| \y)= c^{-\frac{b^n+K+2}{2}}\exp\left\{-\frac{q^{n+1}}{2(q^{n}+1)}\left[\B^n_{kk}+\frac{q^n}{q^n+1}(\y-\Btheta^n_k)^2\right]c^{-1}\right\},
\]
and 
\[
p(A|\y)=  |A|^{-\frac{b^n+K+1}{2}}\exp\left\{-\frac{q^{n+1}}{2q^{n}}tr(\B^n_{-k|k}A^{-1})\right\}.
\]
Notice that $p(A|\y)$, $p(c|\y)$, and $p(a|A,\y)$ are not necessarily the density functions of $A$, $c$ and $a|A$.

Therefore, we have
\begin{equation}\label{eq:kl}
D_{KL}(\B)=\log\xi^0+\mathrm{E}\log\frac{\xi(A)}{p(A|\yx)}+\mathrm{E}\log\frac{\xi(c)}{p(c|\yx)}+\mathrm{E}\log\frac{\xi(a|A)}{p(a|A,\yx)}.
\end{equation}
We now derive the terms in \eqref{eq:kl} one by one.

First, 
\begin{equation}\label{eq:xi0}
\log\xi^0=\frac{b^{n+1}}{2}\log|\B_{-k|k}|+\frac{b^{n+1}}{2}\log\B_{kk}.
\end{equation}

Second, according to the Inverse Wishart distribution of $c$, we have
\[
\mathrm{E}\log c^{-1}\propto \log \B^{-1}_{kk}
\]
and
\[
\mathrm{E}c=(b^{n+1}-K+1)\B_{kk}.
\]
Thus, we obtain
\begin{equation}\label{eq:xic}
\mathrm{E}\log\frac{\xi(c)}{p(c|\yx)}\propto\frac{b^n-b^{n+1}+1}{2}\log\B_{kk}+\frac{q^{n+1}(b^{n+1}-K+1)\B^n_{kk}}{2(q^{n}+1)\B_{kk}}\left[1+\frac{q^n(\y-\Btheta^n_k)^2}{(q^n+1)\B^n_{kk}}\right].
\end{equation}

Third, according to the Inverse Wishart distribution of $A$, we have 
\[
\mathrm{E}\log |A|\propto\log |\B_{-k|k}|
\]
and
\[
\mathrm{E}A^{-1}=b^{n+1}\B^{-1}_{-k|k}.
\]
Thus, we obtain that
\begin{equation}\label{eq:xia}
\mathrm{E}\log\frac{\xi(A)}{p(A|\yx)}\propto \frac{b^n-b^{n+1}}{2}\log |\B_{-k|k}|+\frac{q^{n+1}b^{n+1}}{2q^n}\mathrm{tr}(\B^n_{-k|k}\B^{-1}_{-k|k}).
\end{equation}

Lastly, according to the multivariate normal distribution of $a|A$, we have
\begin{equation}\label{eq:aA}
\mathrm{E}\log\frac{\xi(a|A)}{p(a|A,\yx)}\propto \left(\frac{\B_{-k,k}}{\B_{kk}}-\frac{\B^n_{-k,k}}{\B^n_{kk}}\right)^\top \B^{-1}_{-k|k}\left(\frac{\B_{-k,k}}{\B_{kk}}-\frac{\B^n_{-k,k}}{\B^n_{kk}}\right).
\end{equation}
Combine \eqref{eq:xi0}--\eqref{eq:aA}, the objective function can be expressed as
\begin{subequations}
\begin{align}
D_{KL}(\B) & =\frac{b^n+1}{2}\log\B_{kk}+\frac{q^{n+1}(b^{n+1}-K+1)\B^n_{kk}}{2(q^{n}+1)\B_{kk}}\left[1+\frac{q^n(\y-\Btheta^n_k)^2}{(q^n+1)\B^n_{kk}}\right] \label{DKL-1}
\\ 
 & +\frac{b^n}{2}\log |\B_{-k|k}|+\frac{q^{n+1}b^{n+1}}{2q^n}\mathrm{tr}(\B^n_{-k|k}\B^{-1}_{-k|k}) \label{DKL-2}\\ 
 & +\frac{q^{n+1}\B^n_{kk}}{2q^n}\left(\frac{\B_{-k,k}}{\B_{kk}}-\frac{\B^n_{-k,k}}{\B^n_{kk}}\right)^\top \B^{-1}_{-k|k}\left(\frac{\B_{-k,k}}{\B_{kk}}-\frac{\B^n_{-k,k}}{\B^n_{kk}}\right). \label{DKL-3}
\end{align}
\end{subequations}

We next minimize $D_{KL}(\B)$ with respect to $\B$. It is clear that this can be done by minimizing $D_{KL}(\B)$ with respect to $\B_{kk}$, $\frac{\B_{-k,k}}{\B_{kk}}$, and $\B_{-k|k}$. We first observe that only \eqref{DKL-3} involves term $\frac{\B_{-k,k}}{\B_{kk}}$. For any fixed $\B_{kk}$ and $B_{-k|k}$, the minimizer of \eqref{DKL-3} is given by $(\frac{\B_{-k,k}}{\B_{kk}})^* = \frac{\B^n_{-k,k}}{\B^n_{kk}}$, and the corresponding minimum of \eqref{DKL-3} is $0$. We then notice that \eqref{DKL-1} only involves $\B_{kk}$, and \eqref{DKL-2} only involves $\B_{-k|k}$, by optimizing \eqref{DKL-1} and \eqref{DKL-2} with respect to $\B_{kk}$ and $\B_{-k|k}$, respectively, we get:
\[
(\B_{-k|k})^*=\frac{b^{n+1}q^{n+1}}{b^{n}q^{n}}\B^n_{-k|k},
\]
\[
(\B_{k,k})^*=\frac{q^{n+1}(b^{n+1}-K+1)\left[\B^n_{k,k}+\frac{q^n}{q^n+1}(\y-\Btheta^n_k)^2\right]}{(b^n+1)(q^{n}+1)}.
\]
\end{proof}

\end{document}